\documentclass[aps,prl,nobibnotes,twocolumn,showkeys,superscriptaddress]{revtex4-1}

\usepackage[artemisia]{textgreek}
\usepackage{amssymb}
\usepackage{epsfig}
\usepackage{amsmath}
\usepackage{graphicx}
\usepackage{dcolumn}
\usepackage{latexsym}
\usepackage{color}
\usepackage{epstopdf}
\usepackage{subfigure}
\usepackage{nicefrac}
\usepackage{blindtext}
\usepackage[ulem=normalem]{changes}
\usepackage{float}
\usepackage{xcolor}
\usepackage{soul}
\usepackage[T1]{fontenc}
\usepackage[hidelinks]{hyperref}
\usepackage{xcolor}
\setcitestyle{super}
\hypersetup{
    colorlinks,
    linkcolor={red!50!black},
    citecolor={blue!50!black},
    urlcolor={blue!80!black}
}
\usepackage{lipsum}

\usepackage[symbol]{footmisc}

\begin{document}


\title{Epitaxially Driven Phase Selectivity of Sn in Hybrid Quantum Nanowires}

\author{Sabbir A. Khan$^\clubsuit$} 
\email{skh@dfm.dk}
\altaffiliation{$^\clubsuit$Contributed equally to this work}
\affiliation{Center for Quantum Devices, Niels Bohr Institute, University of Copenhagen, 2100 Copenhagen, Denmark}
\affiliation{Danish Fundamental Metrology, Kogle Alle 5, 2970 Hørsholm, Denmark}

\author{Sara Mart\'{i}-S\'{a}nchez$^\clubsuit$} 
\email{sara.marti@icn2.cat}
\affiliation{Catalan Institute of Nanoscience and Nanotechnology (ICN2), CSIC and BIST, Campus UAB, Bellaterra, 08193 Barcelona, Catalonia, Spain}

\author{Dags Olsteins}
\affiliation{Center for Quantum Devices, Niels Bohr Institute, University of Copenhagen, 2100 Copenhagen, Denmark}

\author{Charalampos Lampadaris}
\affiliation{Center for Quantum Devices, Niels Bohr Institute, University of Copenhagen, 2100 Copenhagen, Denmark}

\author{Damon James Carrad}
\affiliation{Center for Quantum Devices, Niels Bohr Institute, University of Copenhagen, 2100 Copenhagen, Denmark}
\affiliation{Department of Energy Conversion and Storage, Technical University of Denmark, Fysikvej, Building, Lyngby, 310, 2800 Denmark}

\author{Yu Liu}
\affiliation{Center for Quantum Devices, Niels Bohr Institute, University of Copenhagen, 2100 Copenhagen, Denmark}

\author{Judith Qui\~{n}ones}
\affiliation{Catalan Institute of Nanoscience and Nanotechnology (ICN2), CSIC and BIST, Campus UAB, Bellaterra, 08193 Barcelona, Catalonia, Spain}

\author{Maria Chiara Spadaro}
\affiliation{Catalan Institute of Nanoscience and Nanotechnology (ICN2), CSIC and BIST, Campus UAB, Bellaterra, 08193 Barcelona, Catalonia, Spain}

\author{Thomas S. Jespersen}
\affiliation{Center for Quantum Devices, Niels Bohr Institute, University of Copenhagen, 2100 Copenhagen, Denmark}
\affiliation{Department of Energy Conversion and Storage, Technical University of Denmark, Fysikvej, Building, Lyngby, 310, 2800 Denmark}

\author{Peter Krogstrup}
\email{krogstrup@nbi.dk}
\affiliation{NNF Quantum Computing Programme, Niels Bohr Institute, University of Copenhagen, Denmark}

\author{Jordi Arbiol}
\email{arbiol@icrea.cat}
\affiliation{Catalan Institute of Nanoscience and Nanotechnology (ICN2), CSIC and BIST, Campus UAB, Bellaterra, 08193 Barcelona, Catalonia, Spain}
\affiliation{ICREA, Pg. Lluís Companys 23, 08010 Barcelona, Catalonia, Spain}

\date{\today}

\begin{abstract}

Hybrid semiconductor/superconductor nanowires constitute a pervasive platform for studying gate-tunable superconductivity and the emergence of topological behavior. Their low-dimensionality and crystal structure flexibility facilitate novel heterostructure growth and efficient material optimization; crucial prerequisites for accurately constructing complex multi-component quantum materials. Here, we present an extensive optimization of Sn growth on InSb, InAsSb and InAs nanowires. We demonstrate how the growth conditions and the crystal structure/symmetry of the semiconductor drive the formation of either semi-metallic $\mathrm{\alpha-Sn}$ or superconducting $\mathrm{\beta-Sn}$. For InAs nanowires, we obtain phase-pure, superconducting $\mathrm{\beta-Sn}$ shells. However, for InSb and InAsSb nanowires, an initial epitaxial $\mathrm{\alpha-Sn}$ phase evolves into a polycrystalline shell of coexisting $\mathrm{\alpha}$ and $\mathrm{\beta}$ phases, where the $\beta/\alpha$ volume ratio increases with Sn shell thickness. Whether these nanowires exhibit superconductivity or not critically relies on the $\mathrm{\beta-Sn}$ content. Therefore, this work provides key insights into Sn phase control on a variety of semiconductors, with consequences for the yield of superconducting hybrids suitable for generating topological systems.

\end{abstract}


\maketitle


\begin{figure*}[ht!]
\includegraphics[width = 0.95\textwidth]{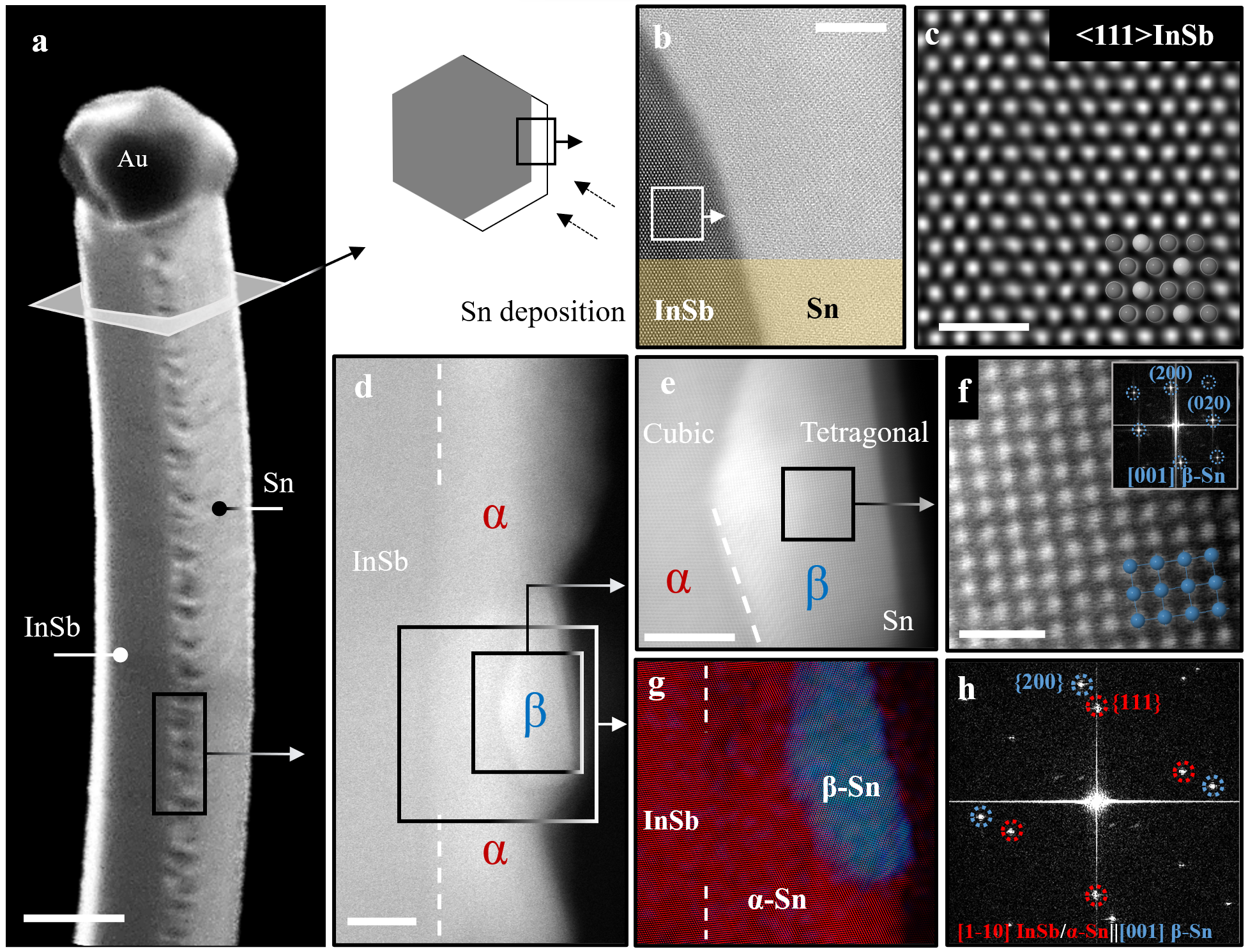}
\caption{{\textbf{Hybrid InSb-Sn nanowires with multiple Sn phases.}} \textbf{a}, Scanning electron microscopy (SEM) image of a MBE grown InSb nanowire \textit{in situ} hybridized with a Sn shell. \textbf{b}, Cross-sectional aberration-corrected HAADF-STEM micrograph of the InSb-Sn interface. \textbf{c}, Magnified HAADF-STEM image from the InSb segment showing cubic zinc blende crystal structure. \textbf{d}, Low magnification longitudinal HAADF-STEM micrograph of the heterostructure where coexistence of both $\mathrm{\alpha-Sn}$ and $\mathrm{\beta-Sn}$ is observed. Dashed line indicates the starting of the Sn shell. \textbf{e}, Magnified image of the interface between cubic-$\mathrm{\alpha}$ and tetragonal-$\mathrm{\beta}$ phases. The white dashed line marks a $\{$111$\}$ $\mathrm{\alpha-Sn}$ plane interfacing with a $\mathrm{\beta-Sn}$ domain. \textbf{f}, Zooming of the $\mathrm{\beta-Sn}$. \textbf{g}, Fourier filtered map of crystal planes from the squared region marked in panel (d). Blue section matches $\mathrm{\beta-Sn}$ phase while red corresponds to InSb and $\mathrm{\alpha}$ phases which could not be unambiguously decoupled due to the equivalent cubic symmetries and their remarkably low lattice mismatch. \textbf{h}, Fast Fourier transform employed for (g) structural map showing undistinguishable plane reflections for InSb and $\mathrm{\alpha-Sn}$ (red circles) while blue circles correspond to \{200\} planes of [001] oriented $\mathrm{\beta-Sn}$. Scale bars are: (a) 100 nm, (b) 5 nm, (c) 1 nm, (d) 20 nm, (e) 10 nm, (f) 1 nm.}
\label{fig1}
\end{figure*}

Tin is a common group IV element used in a broad range of industrial applications including electronic circuits, optoelectronic devices, energy storage devices and coating for commercial products\cite{dean1987tinplate,ng2013unveiling, kufner2013structural, kufner2013optical, black2005getting, hormann2015semiconductor, tu2018fast}. Recently, interest was further triggered with the promise of exploring exotic topological phases with Sn-semiconductor hybrids \cite{xu2017elemental, ohtsubo2013dirac, pendharkar2021parity, khan2020highly}. However, the type of topology depends on which of the two major allotropes ($\mathrm{\alpha-Sn}$ or $\mathrm{\beta-Sn}$) are present. Hybridising the semi-metallic, cubic $\mathrm{\alpha-Sn}$ phase\cite{busch1960semiconducting, ewald1958gray, groves1963band, farrow1981growth, legrain2016understanding, didschuns2002superconductivity, molodets2000thermodynamic, smith1985alpha, de1935electrical, hormann2015semiconductor, didschuns2002superconductivity, molodets2000thermodynamic}, with semiconductors such as InSb (111) or (100) induces lattice mismatch-related strain fields. The resulting broken cubic symmetry can lead to topological insulator behaviour\cite{xu2017elemental,song2019thermal,roman1972stress,osaka1994surface, fu2007topological,groves1963band,barfuss2013elemental,ohtsubo2013dirac, zheng2019epitaxial}. Conversely, metallic $\mathrm{\beta-Sn}$ is a comparatively dense body-centered tetragonal structure, and exhibits superconductivity with a bulk critical temperature of 3.7 K. \cite{busch1960semiconducting, ewald1958gray, groves1963band, farrow1981growth, legrain2016understanding, didschuns2002superconductivity, molodets2000thermodynamic, smith1985alpha, de1935electrical, hormann2015semiconductor, didschuns2002superconductivity, molodets2000thermodynamic}. Hence, hybrids of $\mathrm{\beta-Sn}$ and one-dimensional semiconductors with strong spin-orbit interaction -- such as InSb or InAs -- may exhibit topological superconductivity in the presence of magnetic fields \cite{Majorana1937, kitaev2001, Lutchyn2010, oreg2010helical, Mourik2012, Rokhinson2012, Deng2016, Das2012, Lutchyn2018, frolov2020topological, leijnse2012introduction, prada2020andreev}.

Unleashing the potential of either of the two Sn phases --and excluding the potential for unwanted behaviours-- requires phase pure crystal growth, which is a challenge in thin films\cite{liu2022growth, didschuns2002superconductivity, hochst1985microscopic, oehl2015critical}. In particular, the presence of $\mathrm{\alpha-Sn}$ significantly reduces the yield of superconducting devices\cite{didschuns2002superconductivity} and phase-mixed materials present problems with regard to fabrication. While research has been undertaken on crystal phase stability in bulk Sn,\cite{burgers1957mechanism, musgrave1963relation, ravelo1997equilibrium, yomogita1972geometry, legrain2016understanding, smith1985alpha, hormann2015semiconductor, liu2022growth, didschuns2002superconductivity, hochst1985microscopic, ojima1993direct, oehl2015critical, song2019thermal} a comprehensive study of nanoscale Sn phase formation on semiconductor nanowires (NWs) is yet to be performed. Such knowledge is crucial to determine a selective phase growth and control the interface properties of the Sn-based hybrid quantum heterostructures.



In this article, we perform Sn thin-film growth on different semiconductor NWs and present an in-depth analysis. We tune the growth parameters to determine the optimal conditions and investigate associated structural changes in the Sn film. When grown on B-polar cubic InSb and InAsSb NWs\cite{khan2020highly, mata2012, mata2019}, Sn grows predominantly in the lattice matched $\mathrm{\alpha-phase}$. With increasing film thickness, we observe an increasing density of embedded $\mathrm{\beta-phase}$ grains. Based on the comprehensive crystal analysis of the growth evolution, we suggest a nucleation mechanism for $\mathrm{\beta-Sn}$ embedded within an $\mathrm{\alpha-Sn}$ matrix. Obtaining hybrids with phase-pure $\mathrm{\beta-Sn}$ was possible by using InAs semiconductor NW. Here, the $\mathrm{\beta-Sn}$ creates crystal domain matched grains, while differences in crystal symmetry between InAs and $\mathrm{\alpha-Sn}$ and the large associated plane mismatches strongly suppress $\mathrm{\alpha-Sn}$ nucleation. Finally, we correlate structural properties with electrical behaviour at low-temperatures. We show a strong correlation between the presence of superconductivity in ZB (InSb/InAsSb)-Sn with the $\mathrm{\beta}$ grain density, and that those films expected to have a vanishingly low $\mathrm{\beta-Sn}$ content exhibit no trace of superconductivity. Conversely, $\mathrm{\beta}$ dominant WZ(InAs)-Sn based hybrids exhibit an induced superconducting gap with energy $\Delta = 440 \mu$eV.

\section{Results and Discussion}
\textbf{Multiphase Sn crystals in hybrid NWs.} Au catalyst-assisted NWs were grown on InAs (111)B or InAs (100) \cite{khan2020highly} substrates using a molecular beam epitaxy (MBE) system. Similar to our earlier works \cite{khan2020highly,khan2021multiterminal,carrad2022photon,stampfer2022andreev}, InSb and ternary InAsSb NWs were grown from InAs stems. Once the NWs were grown, Sn deposition was performed in a high vacuum physical vapor deposition (PVD) chamber, which is $\textit{in vacuo}$ connected to the MBE (see Methods for details). Fig. \ref{fig1} (a) shows a scanning electron microscopy (SEM) image of an InSb NW hybridized with approximately 20 nm Sn shell. Here, Sn was grown on the selected facets of the hexagonal NW as indicated in the image. The crystal parameters and the elemental distribution mapping of these NWs are shown in Supporting Information S1 and S2.


For crystal analysis, aberration-corrected high angular annular dark field (HAADF) - scanning transmission electron microscopy (STEM) micrographs were acquired in a cross-section of the hybrid NWs (Fig.\ref{fig1} (b)). Clear contrast between InSb core (dark) and Sn shell (bright) can be observed. The InSb segment can be identified as a cubic zinc blende (ZB) crystal with a lattice constant of 6.479 Å (see magnified HAADF-STEM micrograph in Fig.\ref{fig1} (c)). In Fig.\ref{fig1} (b), Sn seems to form a single crystal, but unambiguous phase identification was not possible either due to the $\mathrm{\alpha-\beta}$ overlap or elemental migration and damage induced during focused ion beam (FIB) assisted cross-section preparation (see Supporting Information S3). Therefore, we performed structural analysis on longitudinal views of NWs directly transferred to Cu grids by employing a combination of High Resolution TEM (HRTEM) and aberration corrected HAADF-STEM imaging (see Fig.\ref{fig1} (d)).  Here three different HAADF intensities can be observed: the darker contrast originated from the InSb core, the deposited Sn shows the light contrast with an enclosed area in the shell. Taking a closer look into Fig.\ref{fig1} (d) we can observe that most of the shell matches the $\mathrm{\alpha-Sn}$ phase, which is also a face-centered cubic with a lattice constant of 6.489 Å. The small lattice mismatch ($\epsilon \sim$ 0.15$\%$) between $\mathrm{ZB-InSb}$ and $\mathrm{\alpha-Sn}$ favors its growth on the NW interface, which we will elaborate in the later discussion. The brightest small domain can be identified as a $\mathrm{\beta-Sn}$ grain embedded in the $\mathrm{\alpha-Sn}$ matrix. From this and other analyzed NWs (see Supporting Information S4), we observed that $\mathrm{\beta}$ grains are mostly not grown directly interfacing the InSb NW core, but they seem to appear on the epitaxial cubic $\mathrm{\alpha-Sn}$ shell. The $\mathrm{\beta}$ grains typically show well-defined \{020\} faceting and seem to arise from \{111\} faceting on the $\mathrm{\alpha}$ shell (indicated with white lines in Fig.\ref{fig1} (e)). 
An atomic-resolution [001]-oriented $\mathrm{\beta}$ phase arrangement and its diffraction pattern are shown in Fig.\ref{fig1} (f). Further, in Fig.\ref{fig1} (g) we present a Fourier plane filtered phase map from the atomically resolved micrograph of the hybrid NW to highlight the multi-phase structure. As we can see, due to the quasi identical atomic arrangement and atomic number, InSb core and $\mathrm{\alpha-Sn}$ have a perfect epitaxy and are barely distinguishable. On the other hand, due to the different crystal structure and reduced unit cell volume, the $\mathrm{\beta-Sn}$ grain shows distinct contrast and exhibits a sharp interface with the other cubic structures. These differences in lattice arrangement can be noticed in the fast Fourier transform (FFT) displayed in Fig.\ref{fig1} (h). 

\textbf{Optimal conditions for Sn shell on NWs.} Thin-film formation of Sn crucially depends on the substrate temperature and chemical potential of the incorporation sites, as they exponentially govern the adatom diffusion length on the NW facets \cite{thompson1990epitaxial, krogstrup2015epitaxy, khan2020highly}. Low-temperature growth assists to minimize the adatom diffusion length and drives Sn film to wet on the NW surface. Besides, the chemical potential of the incorporation sites depends on the change in Gibbs free energy that is determined by the interface energy density between NW facets and deposited Sn \cite{thompson1990epitaxial, krogstrup2015epitaxy, khan2020highly}. In the initial stage of the Sn growth, the interface energy depends on the lattice mismatch between two materials and the area of the interface. In the later stage, strain energy also plays a key role in defining the interface energy density and evolving grain boundaries. Hence, selection of NW and a Sn phase with low residual mismatch will ease the formulation of continuous Sn shell. In contrary, if the NW and Sn phase have a high residual mismatch then a significant thermodynamic force will require to make a continuous film on NW facets. Further, a high flux rate also assists Sn shell growth by increasing adatom concentration on the NW facets. High adatom concentration increases the nucleation density on the given surface, which also lowers the average adatom diffusion length. In a nutshell, a combination of the low substrate temperature, favorable interface energy density and high flux rate can provide an optimized Sn shell on the NW facets. Hence, we maintained the Sn flux rate in the range of 3-5 \AA/s and the sample holder temperature was either approximately -100$^{\circ}$C or -150$^{\circ}$C during the growth. A detailed study of growth temperature effect on the hybrid Sn shell structure is presented in the Supporting Information S5-S9.

\begin{figure*}[ht!]
\includegraphics[width=0.93\textwidth]{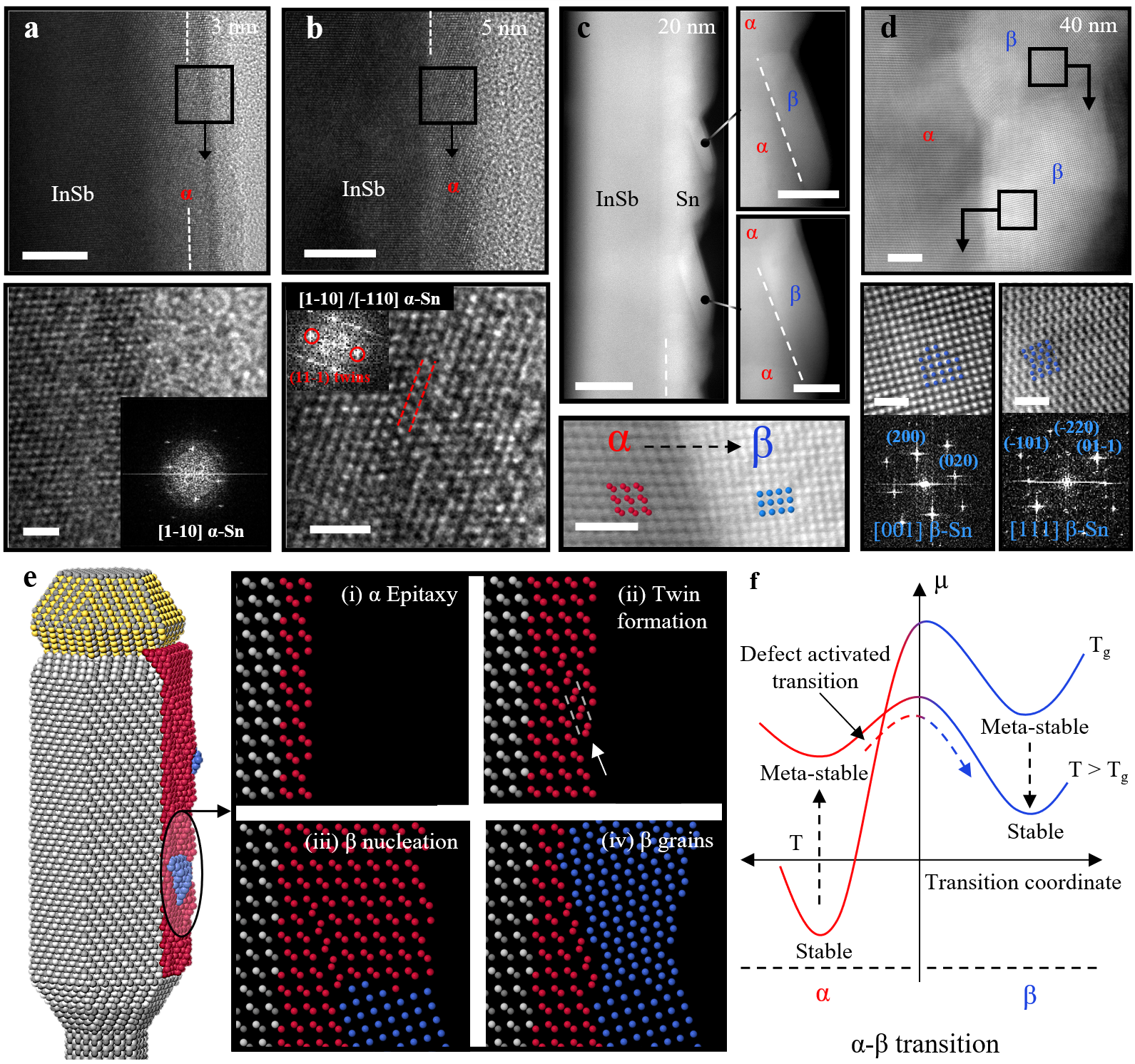}
\caption{{\textbf{Sn shell evolution and $\mathrm{\beta-Sn}$ grain formation.}} \textbf{a}, HRTEM micrograph of the InSb-Sn NW, where approximately 3nm Sn was deposited. Due to the ultra-thin film and epitaxial relation the interface is just distinguished by a subtle contrast variation. FFT on a magnified image of the interface confirms  $\mathrm{\alpha-Sn}$ phase. \textbf{b}, HRTEM image of the approximately 5nm Sn shell where an initial stage twin plane is identified. \textbf{c}, Large-scale low magnification HAADF-STEM micrograph of 20 nm Sn shell where clear contrast between $\mathrm{\alpha-Sn}$ and $\mathrm{\beta-Sn}$ can be observed. Zoom-in shows $\mathrm{\beta-Sn}$ grains with well-defined faceting that seem to arise from $\{$111$\}$ faceting of the $\mathrm{\alpha-Sn}$ shell (white dot lines). The bottom panel shows an atomically resolved HAADF-STEM micrograph of the $\mathrm{\alpha}-\mathrm{\beta}$ interface. \textbf{d}, HAADF-STEM micrograph of 40 nm Sn deposited on InSb. $\mathrm{\beta-Sn}$ grains can be observed on top of initial $\mathrm{\alpha-Sn}$. \textbf{e}, Hypothesis for the $\mathrm{\alpha,\beta}$ nucleation of Sn film on InSb NW. (i) Initially $\mathrm{\alpha-Sn}$ grows epitaxially on the lattice matched InSb cubic interface. (ii) Few layers away from interface twin defects appear (indicated by the arrow) mostly oblique to the growth direction. (iii) These twin planes act as a preferential nucleation centers for the $\mathrm{\beta-Sn}$ phases when activation energy barrier is overcome and hence, $\mathrm{\beta}$ grain forms. (iv) For thick shells, large $\mathrm{\beta-Sn}$ grains with different orientation can be observed and initial \{020\}-$\beta$ faceting dissolves. \textbf{f}, Schematic of the $\mathrm{\alpha\leftrightarrow\beta}$ transition as a function of chemical potential ($\mathrm{\mu}$) and temperature. Here, $\mathrm{T_{g}}$ (during growth) and $\mathrm{T}$ are the holder temperatures. Scale bars are: (a) 10 nm \& 2 nm, (b) 10 nm \& 2 nm, (c) left: 50 nm, right: 20 nm (both), bottom: 2 nm, (d) top: 5 nm, bottom: 1 nm (both).}
\label{fig2}
\end{figure*}

\textbf{$\beta$-Sn nucleation in $\alpha$-Sn matrix.} To understand the thin film evolution we performed an extensive structural study varying the Sn shell thickness (Fig.\ref{fig2} (a-d)). In Fig.\ref{fig2} (a), approximately 3 nm Sn was deposited on an InSb NW. As $\mathrm{\alpha-Sn}$ forms an epitaxy with InSb, it was difficult to differentiate the interface. However, a weak contrast fluctuation from the interface (indicated with the white dot line) can be observed. We did not see any tetragonal $\mathrm{\beta-Sn}$ domain in this extremely thin shell, which would have been noticeable on the cubic structure. More details can be found in Supporting Information S10. In Fig.\ref{fig2} (b), approximately 5 nm of Sn was deposited on the InSb NWs. Here, a dominance of $\mathrm{\alpha-Sn}$ is still observed and some twinning planes start to be visible, as indicated in the magnified section and inset diffraction pattern. In addition, a few seed-stage $\mathrm{\beta-Sn}$ grain domains are also observed in this case (see Supporting Information S11). In some sporadic cases, we observed $\mathrm{\beta-Sn}$ grains close to the interface with InSb, but unambiguous resolution of possible direct contact between InSb and $\mathrm{\beta-Sn}$ could not be determined due to the 3D structure of the system. In any case, the threshold thickness for which we start to observe $\mathrm{\beta-Sn}$ nucleation is 5 nm, so we assume some epitaxial $\mathrm{\alpha-Sn}$ might be present prior to it even though it might not be the only $\mathrm{\beta-Sn}$ formation mechanism. In Fig.\ref{fig2} (c), approximately 20 nm of Sn was deposited and full-fledged faceted [001]-oriented $\mathrm{\beta-Sn}$ growing on the \{11-1\} $\mathrm{\alpha-Sn}$ planes can be observed. 
In the magnified inset images, a clear preferential faceting of $\mathrm{\beta-Sn}$ on $\mathrm{\alpha}$ can be observed, which belongs to the \{020\}-$\beta$ family of planes (see more examples in Supporting Information S4). At the bottom of Fig.\ref{fig2} (c), we present an atomically resolved micrograph of the $\mathrm{\alpha-\beta}$ interface between cubic and tetragonal Sn phases. Finally, approximately 40 nm deposited Sn shell is shown in Fig.\ref{fig2} (d). The shell exhibits continuous but rough morphology (also see Supporting Information S12). Here all the NWs investigated by TEM showed similar nanostructure, with multiple $\mathrm{\beta-Sn}$ grains being randomly oriented (most of them out-of-axis) and they are grown from $\mathrm{\alpha}$-phase. Fig.\ref{fig2} (d) shows an example of two neighboring $\mathrm{\beta-Sn}$ grains oriented in two different zone axes, highlighting the polycrystalline nature of $\mathrm{\beta-Sn}$ shell in contrast to the fully epitaxial $\mathrm{\alpha-Sn}$. It is also notable that the strong \{020\}-$\beta$ faceting vanishes as $\beta$ grains grow.  We also observe an increasing density of $\mathrm{\beta-Sn}$ grains presenting different orientations as we increase the shell thickness. Therefore, from a fully epitaxial $\mathrm{\alpha-Sn}$ shell promoted by lattice-matched InSb templates, we observe an evolution to form polycrystalline $\mathrm{\beta-Sn}$ shell in thick films. 

Based on the experimental analysis, in Fig.\ref{fig2} (e) we present a hypothetical model of the structural evolution of Sn phases in the shell grown on a cubic NW template. The chemical potential of the $\mathrm{\alpha-Sn}$ or $\mathrm{\beta-Sn}$ phase on the NW can be associated with the chemical potential of the reference bulk phase ($\mu_{i,bulk}$) and thin film ($\mu_{i, ex}$). The chemical potential in the thin film will again depend on the NW surface ($\mu_{i,surf}$), interface ($\mu_{i,int}$), strain ($\mu_{i,strain}$) and grain boundary ($\mu_{i,gb}$). Hence, $\mathrm{\alpha}$ or $\mathrm{\beta}$ crystal formation is an accumulated factor of all those parameters:

\begin{gather*} 
\mu_{i}=\mu_{i,bulk} +\delta \mu_{i, ex}\\
\delta \mu_{i, ex}=\delta \mu_{i,surf}+\delta \mu_{i,int}+\delta \mu_{i,strain}+\delta \mu_{i,gb} \\ 
\end{gather*}
where, i = $\mathrm{\alpha}$, $\mathrm{\beta}$

Since bulk $\mathrm{\alpha-Sn}$ is stable below 13.2$^{\circ}$C and the interface energy between InSb and $\mathrm{\alpha-Sn}$ is very small due to its close lattice matching, in our low-temperature growth condition, it is favorable to form $\mathrm{\alpha-Sn}$ in the interface. Hence, epitaxial Sn film appears initially (Fig.\ref{fig2} (e)(i)). In the later stage (Fig.\ref{fig2} (e)(ii)), twin formation in $\mathrm{\alpha-Sn}$ shell initiates mostly in the \{11-1\} plane away from the interface. Such twins are oblique to the growth directions and can be found all over the shell as shown in the Supporting Information S5-S9. Although epitaxial $\mathrm{\alpha-Sn}$ is grown at the interface, it is meta-stable in temperature exposure and can change the crystallinity if the chemical potential of $\mathrm{\alpha-Sn}$ overcomes the kinetic energy barrier between $\mathrm{\alpha-\beta}$ phases. The schematic of activation energy dependent $\mathrm{\alpha\leftrightarrow\beta}$ transition is shown in Fig.\ref{fig2} (f) and can expressed as follows:    
 
\begin{gather*}
\Delta\Gamma_{\alpha \beta}=\Gamma_{\alpha \beta}-\Gamma_{\beta \alpha}\\
\Gamma_{\alpha \beta, Sn} \sim \exp \left(-\frac{\delta E_{\alpha \beta}}{k_{B} T}\right)
\end{gather*}

Here, $\Delta\Gamma_{\alpha \beta}$ is the growth rate and $\delta E_{\alpha \beta}$ is the activation energy which is dependent on the $\mu_{i}$. Next, tetragonal $\mathrm{\beta}$ grains nucleate on top of cubic $\mathrm{\alpha-Sn}$ shell overcoming the activation energy barrier. This activation energy barrier may be overcome due to the change in local conditions during the growth causing a transition from $\mathrm{\alpha}$ to $\mathrm{\beta}$ phase and vice-versa. However, a likely possibility is that $\mathrm{\beta}$ nucleation occurs while the sample is abruptly exposed to a temperature which is higher than the growth temperature (see Fig.\ref{fig2} (f)), (perhaps right after the low-T Sn growth). As we observed these $\mathrm{\beta-Sn}$ grains typically display well-defined faceting that seems to arise from $\{$111$\}$ faceting on the $\mathrm{\alpha-Sn}$ (Fig.\ref{fig2} (e)(iii)). Hence, it can be inferred that $\mathrm{\beta-Sn}$ overgrows from $\mathrm{\alpha}$ structure by nucleating in twin-induced rough sites instead of a flat surface of InSb NW. Finally, for thick Sn film (Fig.\ref{fig2} (e)(iv)), $\mathrm{\beta-Sn}$ grains gain volume and often multiple grains coalesce into one large grain and simultaneously more $\mathrm{\beta}$ grains start to nucleate in different places of the shell. As a result, different orientation of $\mathrm{\beta-Sn}$ grains can be observed. Once both crystal phases are stabilised at room temperature, no further transition between $\mathrm{\alpha-Sn}$ and $\mathrm{\beta-Sn}$ happens as the activation barrier for the transition is very high at this point. Hence, both $\mathrm{\alpha}$ and $\mathrm{\beta}$ crystal phases co-exist in the Sn shell on cubic NWs. In Supporting Information S13 we performed thermal annealing on the 20 nm shell NWs and observed no changes in the Sn heterostructure even turning the conditions to completely favor $\mathrm{\beta-Sn}$ phase, proving the structural stability of the achieved configuration.

\begin{figure}[t!]
\includegraphics[scale=0.9]{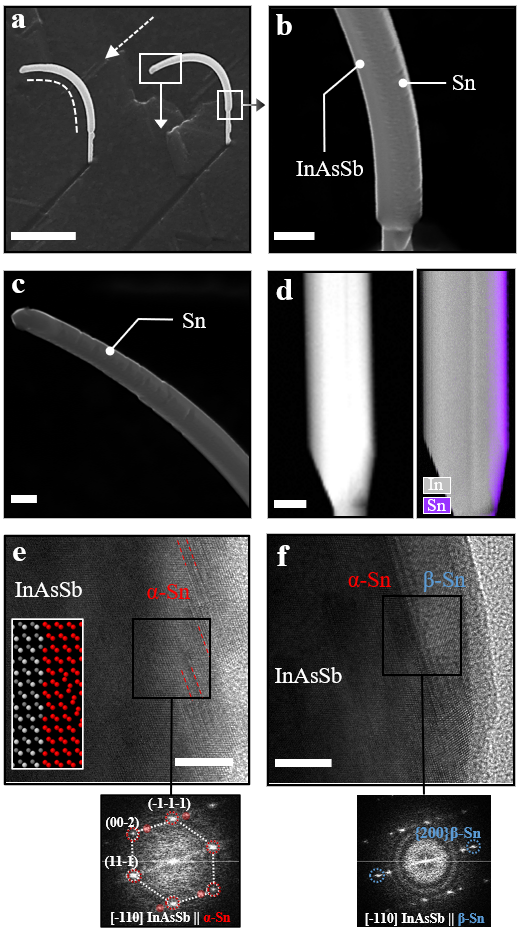}
\caption{{\textbf{Morphology and structural analysis of InAsSb/Sn NWs.}} \textbf{a}, Tilted SEM image of the InAsSb/Sn NWs. Here 20 nm Sn is grown on the three facets of InAsSb NWs. The white arrow shows the deposition direction of the Sn. \textbf{b}, Magnified bottom section of the NW from panel (a). Continuous but rough morphology of the Sn shell can be observed. \textbf{c}, Magnified top section of the NW from panel (a). \textbf{d}, HAADF-STEM image of InAsSb (left) and EELS-based In/Sn compositional map, which reveals continuous Sn coverage. \textbf{e}, HRTEM image and its FFT of $\mathrm{\alpha-Sn}$ on InAsSb. The interface is barely distinguishable due to the same crystal symmetry (see inset model) between cubic $\mathrm{\alpha-Sn}$ and ZB InAsSb, but twinning is present in Sn. \textbf{f}, HRTEM image and its FFT  of observed $\mathrm{\beta-Sn}$ grain on $\mathrm{\alpha-Sn}$ on InAsSb. Scale bars are: (a) 1 {\textmu}m, (b-d) 100 nm, (e-f) 20 nm.}
\label{fig3}
\end{figure}


\textbf{Increasing interfacial lattice mismatch.} Based on the above discussion, it can be presumed that the lattice parameter of the NW core is key to control the Sn-shell phase formation in the hybrid. Therefore, other NW compositions were employed to detriment the lattice match of $\mathrm{\alpha-Sn}$ and hence, hinder its formation. With this purpose, we grew NWs with 80$\%$ Sb and 20$\%$ As composition which implies a lower lattice constant than in pristine InSb NWs given Vegard's law \cite{vegard1921konstitution}, while preserving ZB cubic phase. This configuration implies an increase in lattice mismatch between cubic $\mathrm{\alpha-Sn}$ and the NWs, rising from 0.15 \% with pure InSb to 1.5 \% for InAs$_{0.2}$Sb$_{0.8}$. 

Fig.\ref{fig3} (a) presents InAsSb NWs with a 20 nm Sn shell grown on selected facets of the NWs, where the growth rate was approximately 3{\AA}/s and the sample holder temperature was approximately -150$^{\circ}$C. Magnified sections of these NWs are shown in panel (b) (bottom part) and in panel (c) (top part). Two major observations from these hybrids are: First, Sn shell exhibits a continuous but inhomogeneous morphology on InAsSb NWs compared to InSb NWs; Second, the hybrid InAsSb-Sn NWs show strong bending along the deposition direction. Fig.\ref{fig3} (d) presents a low magnified HAADF-STEM image and an Electron Energy Loss Spectroscopy (EELS) elemental map of In/Sn in a NW, which confirms the continuity of the shell. Taking a close look into the crystal arrangement of the shell, Fig.\ref{fig3} (e) shows an example of the most commonly observed crystal phase of Sn on the InAsSb NWs, belonging to $\mathrm{\alpha-Sn}$ phase and forming a similar structural configuration as observed in InSb NWs. As previously discussed, the $\mathrm{\alpha-Sn}$ shell contains several planar defects including twinning and stacking faults which form $\sim$ 71$^{\circ}$ with the growth direction plane. The model in the inset of Fig.\ref{fig3} (e) shows the atomic arrangement between $\mathrm{\alpha-Sn}$ and $\mathrm{InAs_{0.2}Sb_{0.8}}$. In addition, we could observe some $\mathrm{\beta-Sn}$ grains embedded in the $\mathrm{\alpha-Sn}$, also exhibiting the same crystal configuration with well-defined faceting in the \{200\} planes, but they are discretely distributed through the NW shell, which displays a prevalence of $\mathrm{\alpha}$ phase. More micrographs on different areas are presented in the Supporting Information (S14-S15).

Hence, despite increasing the interfacial lattice mismatch of the NW with the cubic Sn phase, we still obtain a preference for $\mathrm{\alpha-Sn}$ nucleation. We presume that the increase in interface energy results in compressing strain fields that relax towards the outer Sn shell, and thus, strongly bend the NW (Fig.\ref{fig3} (a)). Schematic of compressive strain fields originating the bending is displayed in Supporting Information S14. As a consequence, this configuration still favours $\mathrm{\alpha-Sn}$ nucleation over $\mathrm{\beta-Sn}$. In addition, this bending can be responsible for the rough morphology observed in the shell. Although Sn crystal structure on InAsSb is identical to the InSb one, careful attention should be put on the resulting electronic behavior of the InAsSb-Sn system because of the strain accumulation at the interface, as it is capable of locally modifying its properties \cite{marti2022sub}.

\begin{figure}[t!]
\includegraphics[scale=2.1]{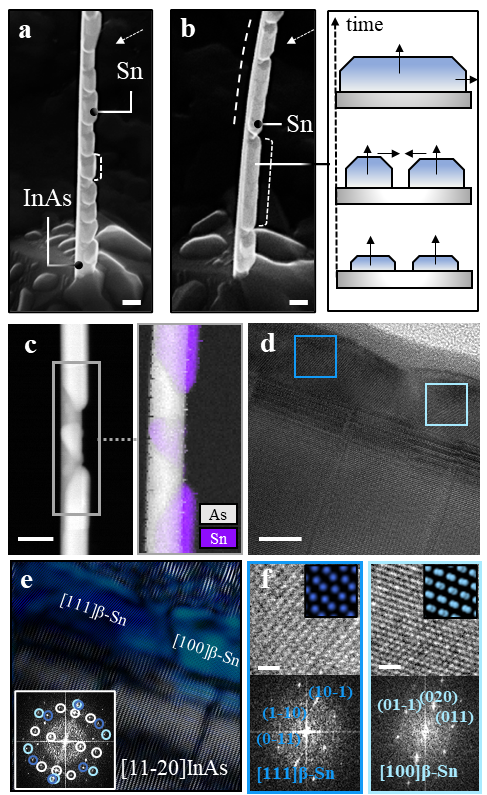}
\caption{{\textbf{Sn shell growth on wurtzite InAs NWs.}} \textbf{a}, Tilted SEM image of the InAs NW with a Sn shell. Sn forms discrete grains with uniform thickness ($\sim$ 22 nm), but non-uniform length (>100 nm). Arrow shows the deposition direction. \textbf{b}, Doubling the Sn deposition time, where Sn shell thickness is $\sim$ 45 nm. With longer growth time nearby Sn grains grow larger and eventually collapse on each other as shown in the right-side schematic. \textbf{c}, HAADF-STEM micrograph of an InAs-Sn NW and EELS compositional mapping of As and Sn on the squared region. \textbf{d}, HRTEM micrograph showing two Sn grains in direct contact with the InAs NW. \textbf{e}, Fourier filtered structural map of (d) showing the spatial distribution of two differently oriented $\mathrm{\beta-Sn}$ grains on InAs. \textbf{f}, Zoom of the squared regions in (d) with their corresponding indexed FFTs and phase identification as $\mathrm{\beta-Sn}$. Scale bars are: (a-c) 100 nm, (d) 10 nm, (f) 1 nm (both).}
\label{fig4}
\end{figure}


\textbf{Selective $\beta$-Sn nucleation in poor lattice matched WZ core.} Finally, we investigate the deposition of Sn on WZ InAs NWs. Fig.\ref{fig4} (a) shows InAs NWs with a $\sim$ 20 nm Sn shell grown at the rate of 3{\AA}/s and holder temperature approximately -150$^{\circ}$C. Sn forms discrete grains with uniform thickness on InAs NW facets. The NW crystallizing in the hexagonal phase implies that there is no particularly favorable lattice matched configuration for any of the two Sn phases, leading to a high interface energy between these materials (see Supporting Information S1) \cite{krogstrup2015epitaxy, khan2020highly}. As a result, a high thermodynamic driving force is needed for Sn to wet on InAs NW facets during the growth. Hence, in our growth conditions with a limited low-temperature range Sn remains as dewetted islands on InAs NW. Therefore, to achieve a continuous film with the same thickness the temperature should be further lowered so that the Sn adatoms can be immobilized during the growth. Another way to achieve a continuous film is by increasing Sn thickness. Fig.\ref{fig4} (b) shows an example of  $\sim$ 45 nm Sn deposited on a InAs NW keeping the same growth conditions. Doubling the Sn thickness enhanced the length of the islands. However, the shell remains discrete and more material should be deposited to achieve a continuous film. As presented in the schematics on the left, upon reaching a critical thickness, the neighboring islands coalesce and evolve to form large islands which can merge again with the adjacent ones. 
Fig.\ref{fig4} (c) shows an EELS-based compositional map of Sn and As of a NW together with a HAADF-STEM micrograph of the area of study. We can observe that despite the grains tend to be relatively flat and homogeneous in terms of thickness, the grain size distribution is highly inhomogeneous. When moving to analyze the crystal phase of Sn, we found most of the grains, including the larger ones, are out-of-axis with respect to the NW (see Supporting Information S16), while we could find some small sections being in-axis with the InAs (see Fig.\ref{fig4} (d-f)). We can identify all these grains as $\mathrm{\beta-Sn}$ (Fig.\ref{fig4} (f)) and they are in direct contact with the InAs NW (Fig.\ref{fig4} (e)), with no presence of $\mathrm{\alpha-Sn}$ prior to their nucleation. More micrographs with $\beta$ phase identification are available in the Supporting Information (S16-S20). Based on the analyzed structures, we presume that the presence of hexagonal crystal as growth template inhibits the formation of the cubic $\mathrm{\alpha-Sn}$ phase. In the previous cases, the same cubic crystal nature of InSb and InAsSb acted as a template for cubic $\mathrm{\alpha-Sn}$ growth and made it the most stable phase by hindering the nucleation of tetragonal $\mathrm{\beta-Sn}$. However, when employing hexagonal InAs as a core material the crystal symmetry equivalence is broken. The most similar configuration between cubic and hexagonal would be InAs(0002)/$\mathrm{\alpha-Sn}$(111)\cite{PhysRevB.80.245325, doi:10.1063/1.2976338}, but this still implies a 7\% plane mismatch along the NW growth axis direction. Therefore, lower mismatch $\mathrm{\beta-Sn}$ growth gets favored instead of $\mathrm{\alpha-Sn}$ and it is the only phase we observed in all the NWs we examined. 
While in previous cases, $\mathrm{\beta-Sn}$ grains nucleated on $\mathrm{\alpha-Sn}$ matrix showed some preferential out-of-plane orientation, $\mathrm{\beta-Sn}$ grains on InAs show a high degree of randomness regarding their orientation with respect to the InAs lattice, especially for smaller grains. However, for the long-range grains, we observe a dominant Sn orientation in the in-plane direction, although it is not in axis with the InAs crystal. In Supporting Information S16-S18 multiple HRTEM micrographs of the InAs-Sn heterostructure with larger Sn grains are presented. A 2.9{\AA} plane reflection parallel to InAs(0002) is visible in all of them and creates periodic Moiré patterns in areas overlapping with InAs. This reflection belongs to the \{020\}$\mathrm{\beta-Sn}$ family of planes. Given the differences in lattice symmetry between $\mathrm{\beta-Sn}$ and WZ InAs, there is no low-index 3D configuration with low associated mismatch in all matching planes, and this explains the appearance of multiple crystal orientations in the heterostructures. In the case of the observed InAs(0002)/$\mathrm{\beta-Sn}$\{200\}, they provide a mismatch of  $\varepsilon${InAs(0002)/$\mathrm{\beta-Sn}$\{200\}}=16\%. However, the domain formation consisting in  5x(002)InAs with 6x\{200\} $\mathrm{\beta-Sn}$ planes, reduces the mismatch to $\varepsilon${5xInAs(0002)/6x$\mathrm{\beta-Sn}$$\{$200$\}$}=0.017\%. In fact, the Moiré periodicity found in the overlapping areas matches with this plane combination (S16-S18). We assume that this low plane mismatch domain is therefore more energetically favorable than the other observed orientations and can be maintained in a longer spatial range, thus being the orientation observed in Sn larger grains. More details and schematics of the plane matching in the formed domains can be found in S16.    

Last but not least, in order to cross-check that the observed preclusion of $\mathrm{\alpha-Sn}$ growth on InAs NWs is caused by its structural symmetry rather than its composition, we radially grew approximately 2 nm shell of InAs covering the InSb NW before depositing Sn (see Supporting Information S21). In these conditions, InAs gets ZB phase and remains completely strained on top of the InSb, adopting its crystal symmetry and lattice constant. We observed that regardless of the presence of InAs covering InSb, Sn crystallizes mainly in the $\alpha$ phase with some $\beta$ grains being embedded on it (see S21). This fact is consistent with the NW crystal structure being one of the key parameters driving Sn phases on hybrid NWs. A schematics of Sn phase dynamics as a function of lattice parameters is presented in Supporting Information S22.

\begin{figure}
\includegraphics[width=8.65cm]{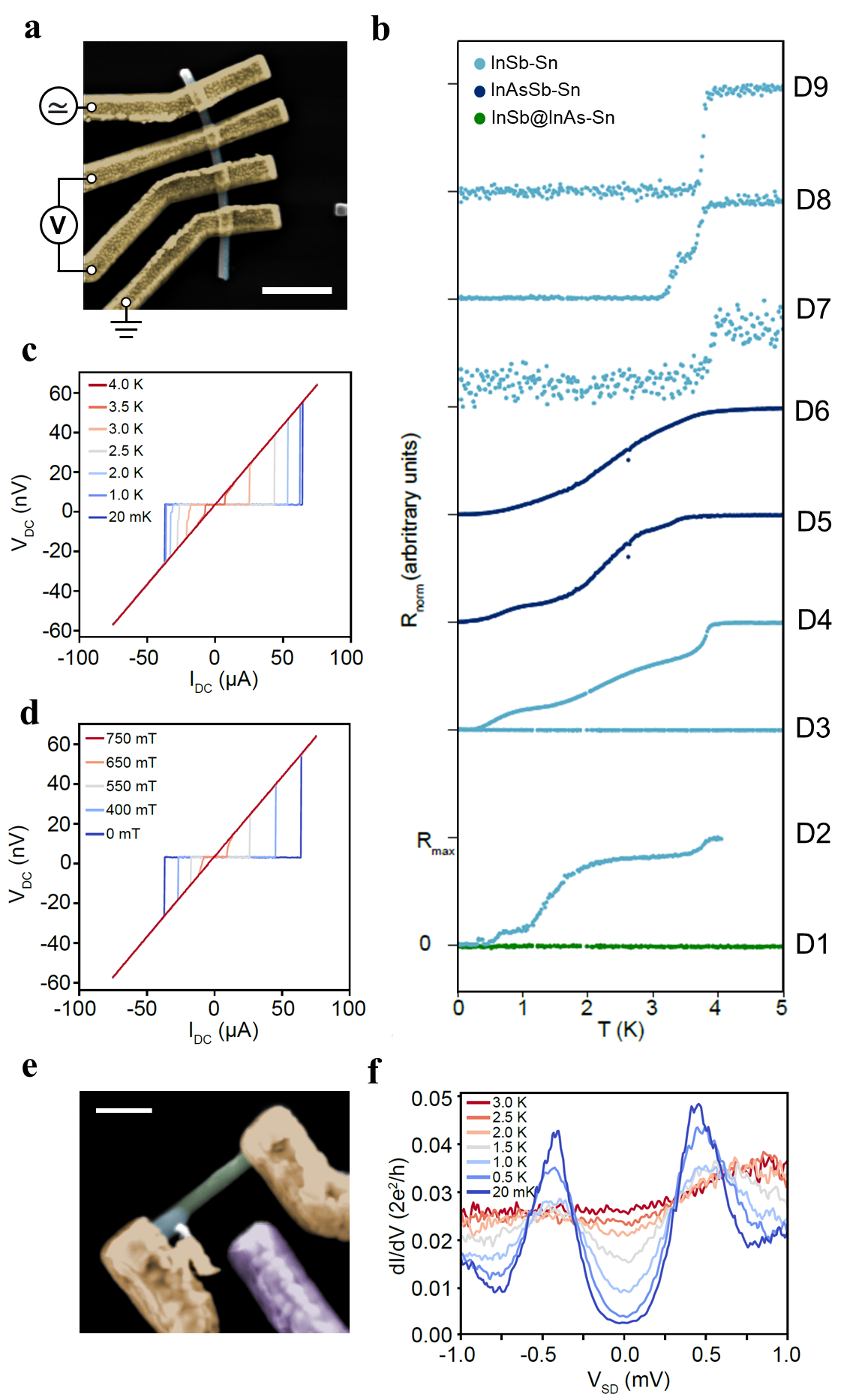}
\vspace{-0.5cm}
\caption{\textbf{Low temperature measurements of NW-Sn hybrid devices.}
\textbf{a}, SEM micrograph of a typical device. \textbf{b}, Temperature dependence of the four-terminal differential resistance for 9 different devices from 3 different hybrid types. See in the text and supporting information S23 for details. For each device the resistance is normalized to the value at the highest temperature and curves are off-set for clarity. \textbf{c,d}, $VI$-curves for device D9 measured as a function of temperature and perpendicular magnetic field respectively. The sweep direction was from negative to positive current. \textbf{e}, SEM micrograph of a two-terminal InAs-Sn N-S device. \textbf{f}, Differential conductance as a function of voltage bias measured for InAs-Sn device in different temperatures. Scale bars are: (a) 1 {\textmu}m, (e), 200 nm.}
\vspace{-0.5cm}
\label{fig5}
\end{figure}

\textbf{Low temperature electrical measurements.} To investigate the superconducting properties of the hybrids, electrical devices were fabricated and measured at low temperatures. Individual hybrid NWs were transferred from the growth substrate to a substrate of highly doped silicon capped with $200 \, \mathrm{nm}$ of SiO$_2$. Nanowires were located with respect to a pre-defined alignment grids and Au/Ti contacts were defined using standard e-beam lithography and metal evaporation. See Methods section for details of the fabrication. Fig. \ref{fig5} (a) shows a SEM micrograph of a typical device with four contacts on the same NW. $10 \, \mathrm{nA}$ \textit{ac}-current was sourced between the outer contacts and the resulting voltage drop between the inner pair was measured using standard lock-in techniques yielding the resistance between the inner pairs. Measurements were performed in a dilution cryostat with a base temperature of $15 \, \mathrm{mK}$. Fig.\ \ref{fig5} (b) shows the temperature dependence of the resistance normalized to the value at the highest temperature for 9 different NWs from 6 different growths of 3 different hybrid types. D2-D4, D7-D9 are InSb-Sn devices, D5-D6 are InAsSb-Sn devices and D1 is InSb(InAs shell)-Sn device. See the table in Supporting Information S23 for the growth parameter details. Devices D3 and D1 show no changes in the resistance upon cooling which we attribute to the absence $\mathrm{\beta-Sn}$ grains in the gap between the inner voltage probes for these devices. All other devices, however, show a clear drop in resistance around $3.7-3.9 \, \mathrm{K}$ close to the the expected transition of $\mathrm{\beta-Sn}$. While D9 shows an abrupt drop to a $R=0 \, \Omega$ at $\sim 3.8 \, \mathrm K$ the remaining devices exhibit a gradual transition with multiple distinguishable steps upon cooling. This behavior is consistent with a Sn shell composed of discrete $\mathrm{\beta-Sn}$ grains electrically coupled through a non-superconducting $\mathrm{\alpha-Sn}$ matrix and the semiconductor NW. At $T^\beta_\mathrm C$, isolated $\mathrm{\beta-Sn}$ grains turn superconducting and while the resistance drops, the resistance remains finite. Other grains have transition temperatures suppressed either due to finite size effects \cite{lang2005finite} or because of inverse proximity effect from the surrounding matrix. Upon lowering the temperature a gradual transition thus occurs until a global superconducting phase is established possibly including proximity coupling neighboring grains through the normal matrix or semiconductor wire. A similar trend was shown in systems of designed Josephson arrays of superconducting regions in a normal matrix \cite{han:2014}. This results are consistent with the structural analysis presented in Fig. \ref{fig1}-\ref{fig3}  for ZB InSb and InAsSb NWs based hybrids. Fig. \ref{fig5} (c-d) shows the the $VI$-characteristics of D9 as a function of temperature and perpendicular field respectively. The measurements were performed sweeping the current from negative to positive and the transition from the resistive to the superconducting state upon sweeping $I$ from a large value towards zero occurs at a smaller absolute current than the transition from the superconducting to the dissipative state observed when increasing $I$ from zero. We attribute this asymmetry to Joule heating in the normal state. The critical current can be observed up to 3.5K (for $B_\perp =0$) and $B_\perp = 650 \, \mathrm{mT}$ (for $T=15 \, \mathrm{mK}$) consistent with panel \textbf{b}.

Finally, to measure the size of the superconducting gap and confirm that superconductivity extends from the Sn shell into the semiconductor NW we further fabricated a two terminal device on the InAs-Sn stem as shown in Fig.\ \ref{fig5} (e). Here a large isolated grain could be seen with one contact positioned on the grain and the other on the adjacent Sn-free semiconductor segment. As explained in Fig.\ \ref{fig4}, the discrete Sn grain here is expected to be only superconducting $\mathrm{\beta-Sn}$ phase without $\alpha$ mix. The electron density in the semiconductor can be tuned using the potential on a nearby electrostatic side-gate and close to pinch-off a quantum dot forms in the semiconductor (see Supporting Information S24 for further characterization). In Coulomb blockade the QD act as a tunnel barrier allowing tunnel spectroscopy of the density of states in the semiconductor. Fig. \ref{fig5} (f) shows the measured differential conductance $dI/dV_\mathrm{SD}$ vs.\ $V_\mathrm{SD}$ for different temperatures. Here $V_\mathrm{SD}$ is the applied voltage bias. Superconducting coherence peaks are observed at $V_\mathrm{SD} = \pm \Delta^* /e$ corresponding to an induced gap of $\Delta^* = 440 \, \mu \mathrm{eV}$ close to the value expected for a gap induced by $\mathrm{\beta-Sn}$. 


\section{Conclusion}


In summary, we have presented the hybridization of Sn with semiconductor NWs and performed an in-depth analysis of Sn crystal phase formation in the heterostructure. InSb, InAsSb and InAs NWs were grown using MBE and later $\textit{in-situ}$ hybridized with Sn by determining optimal conditions and attaining the utmost interface quality.
We have provided a comprehensive structural analysis at the atomic scale that reveals the co-existence of the two different Sn crystal phases ($\mathrm{\alpha-Sn}$ and $\mathrm{\beta-Sn}$) when deposited on cubic crystal NWs like InSb or InAsSb. Based on the experimental assessment in different stages of Sn film, we proposed a hypothesis of $\mathrm{\alpha, \beta-Sn}$ phases nucleation in the hybrid. The NWs with cubic lattice provide a template to epitaxially grow $\mathrm{\alpha-Sn}$ due to their quasi-identical crystal structure (space group, atomic positions and lattice constant). Later, $\mathrm{\beta-Sn}$ nucleates preferentially from the defective sites of the film by overcoming a activation energy and eventually isolated grains are observed to be embedded in the dominant $\mathrm{\alpha-Sn}$ phase. The density of $\mathrm{\beta-Sn}$ grains has been shown to increase with shell thickness. On the contrary, selective growth of $\mathrm{\beta-Sn}$ was achieved on InAs NWs, which is assumed to be a result of the poor lattice match between the WZ hexagonal crystal and the cubic $\mathrm{\alpha-}$phase. In agreement with the structural analysis, low-temperature electrical measurements on InSb-Sn and InAsSb-Sn devices manifest the influence of discrete $\mathrm{\beta-Sn}$ grains on surrounding $\mathrm{\alpha-Sn}$ matrix. While one device demonstrated sharp SU transition others show multi-steps transition which is possibly due to the inverse proximity effect from conducting-nonconducting mix structure. Besides, measuring an isolated $\mathrm{\beta-Sn}$ island on InAs NW we demonstrate the superconducting size and prolongation of superconductivity to the semiconducting base. In brief, along with the optimal growth conditions, choosing the right lattice template is the key to realize desired Sn phases. Therefore, this work presents a crucial development to achieve selective Sn phase in hybrid, and thus, to determine the performance of NW-Sn based topological and superconducting quantum devices.



%


\section{Methods}

\subsection{Hybrid Crystal Growth}

\textbf{Molecular Beam Epitaxy.} Vecco Gen II molecular beam epitaxy (MBE) connected to high vacuum electron gun assisted physical vapour deposition (PVD) system was used to grow NW-Sn hybrids. III-V NWs were grown in the MBE chamber, later the samples were transferred into the PVD chamber through the high vacuum tunnel to grow Sn on the selected facets. All the NW batches were grown on InAs (111)B single-side polished substrate. Au was used a growth catalyst and deposited \textit{in situ} in the MBE chamber before NW growth. In the MBE, material cells (III-V and Au) were pre-heated and stabilized before the growth. Optimized cell parameters for NW growth are: In- (865/815)$^{\circ}$C, As- (400/345)$^{\circ}$C, Sb- (680/500)$^{\circ}$C and Au (1100)$^{\circ}$C. While loading InAs (111)B substrate, standard 200$^{\circ}$C for 2 hours baking at load-lock chamber followed by 250$^{\circ}$C for 1 hours degassing at buffer tunnel were carried out confirming clean InAs surface for the growth. Subsequently, the samples were transferred to the MBE growth chamber where annealing was performed for 6 min at 610$^{\circ}$C substrate temperature under high $\mathrm{As_4}$ over-pressure. Next, the substrate was cooled down to 590$^{\circ}$C and Au was deposited for 1 sec. After Au deposition, the substrates were further lowered to growth temperature which is 447$^{\circ}$C and the In flux was introduced in the growth chamber to nucleate the Au catalysts. For InAs NWs 35 min growth was performed. For InSb and InAsSb NWs, Sb flux was introduced after the InAs stem section. Usually the the stem was grown for 6 min. For InSb, the As flux was turned off after the stem, whereas for InAsSb, As flux was compensated according to the desired Sb, As ratio of the NW. The substrate was cooled down to 150$^\circ$C after the NW growth and transferred to the PVD chamber for hybridizing with Sn. 

\textbf{Physical Vapor Deposition.} For low temperature growth of Sn thin film, the substrate cooling was carried out with the liquid nitrogen supply. The lowest holder temperature reach of our system is approximately $\sim -$ 150$^{\circ}$C, which is recorded with the thermocouple connected at back of the substrate holder. Hence, we would like point that, the real temperature on the sample surface may slightly deviate from the reading. Once the desired growth temperature was reached we waited further to completely stabilize the temperature before preparing the chamber for deposition. Subsequently, the substrate was aligned with respect to the Sn source and the deposition angle was corrected to achieve selected facets (two or three) growth. Details about NW facet selection growth and angle corrections is discussed in ref \cite{khan2020highly}. Next, the source Sn materials were heated up with energetic electrons beam and the evaporated Sn flux rate was monitored through quartz-crystal-microbalance (QCM). Once our desired flux rate was achieved, Sn was deposited on the NW facets. The deposition rate was experimented from 0.5 to 5 \AA/s to find out the optimized morphology. After deposition, the hybrid NW sample was kept on the cold holder for sometime and then unloaded through the load-lock. During the unloading process, the NW was initially exposed to oxygen for approximately 15 min and later vented with nitrogen before introducing to the room temperature.          

\subsection {Morphological and Structural Characterization}

\textbf{Scanning electron microscopy.} JEOL JSM-7800F SEM was used to investigate the morphology of the hybrid NW. Standard operating conditions for these samples are: 15KV accelerating voltage, 10-15 mm working distance with secondary electron detector. Depending on the imaging requirements and investigating materials, we also used 30-degree tilted viewing angle. Further, for detail surface morphology investigation of the hybrid facets we tuned to 2-5KV accelerating voltage with 5mm working distance, which helped us gaining information about roughness and residual metals. \\            

\textbf{(Scanning) Transmission Electron Microscopy.} For structural characterization with (S)TEM, NWs were directly transferred to copper grids by a simple dry transfer technique with a clean room wipe. HRTEM micrographs were acquired in a TECNAI F20 microscope operated at 200 kV. Atomically resolved HAADF data was acquired in a probe corrected TITAN microscope operated at 300 kV and a wiener filter was applied to reduce noise in the displayed images. EELS mapping was carried out in a GATAN QUANTUM Spectrometer coupled to a TECNAI F20 microscope operated at 200 kV with approximately 2 eV energy resolution and 1 eV energy dispersion, collecting the range 350-2398 eV of energy loss. Principal Component Analysis (PCA) was applied to the spectrum images to enhance S/N ratio. In M$_{4,5}$, Sn M$_{4,5}$ Sb M$_{4,5}$ and As L$_{2,3}$ were the edges employed for areal density mapping. 2D and 3D atomic models were created using Rhodius software from the University of Cadiz \cite{BERNAL1998135}. For the presented cross-sections, NWs were transferred to a $\mathrm{SiO_x}$ substrate using a \textbf{micromanipulator} where 100 nm tungsten needle was used to selectively pick the NWs and placed parallel to each other. Next, atomic layer deposition system was used to deposit 30 nm of $\mathrm{AlO_x}$ as a protective layer. Focused Ion Beam (Helios 600 dual beam microscope) was used to prepare cross-sectional NW sections.

\subsection {Device Fabrication}

NW-Sn Devices were fabricated on a highly p-doped Si/SiO$_2$ substrate with a pre-defined alignment grid. Individual NWs were transferred from the growth wafer to the device substrate with a mechanical micro-manipulator carrying a 100 nm tungsten needle and imaged with SEM to define their exact location. Contacts were patterned using electron-beam lithography and e-beam evaporation of Ti/Au (5nm/300nm). A6 resist was used for spin coating for 45 sec with 4000 rpm and pumped in a vacuum chamber for 3 hrs. Patterns were developed in 1:3 MIBK: IPA for 45 sec, passivated in IPA for 30 sec and ashed with oxygen plasma for 30 sec. Prior metal evaporation RF Ar ion milling was employed with 7 Watt for 5 mins at 18 mTorr pressure. The deposition rate of Ti and Au was 0.1 and 0.3 nm/sec respectively. Acetone was used for lift-off. For the devices of Fig. \ref{fig5} (b), the distance of the inner probes was defined by the design software at 250 - 500 nm.



\section{Acknowledgement}
The project was supported by Danish Agency for Higher Education and Science, European Union Horizon 2020 research and innovation program under the Marie Sk\l{}odowska-Curie Grant No. 722176 (INDEED), Microsoft Station Q, the European Research Council (ERC) under Grant No. 716655 (HEMs-DAM). Materials growth of this project is performed at NBI Molecular Beam Epitaxy System and authors acknowledge C. B. S\o{}rensen for all the maintenance and keeping system up and running. ICN2 acknowledges funding from Generalitat de Catalunya 2017 SGR 327. This study was supported by MCIN with funding from European Union NextGenerationEU (PRTR-C17.I1) and Generalitat de Catalunya and by “ERDF A way of making Europe”, by the “European Union”. ICN2 is supported by the Severo Ochoa program from Spanish MINECO (Grant No. SEV-2017-0706) and is funded by the CERCA Programme / Generalitat de Catalunya. Authors acknowledge the use of instrumentation as well as the technical advice provided by the National Facility ELECMI ICTS, node "Laboratorio de Microscopías Avanzadas" at University of Zaragoza. MCS has received funding from the post doctoral fellowship Juan de la Cierva Incorporation from MICINN (JCI-2019) and the Severo Ochoa programme. We acknowledge support from CSIC Interdisciplinary Thematic Platform (PTI+) on Quantum Technologies (PTI-QTEP+).

\section{Author Contributions}
 
 S.A.K conceived the idea, experimental plans, performed substrate fabrication, materials growth and morphological characterization. S.M.S performed the (S)TEM related experiments, analyzed TEM and EELS data and prepared the atomic models. S.A.K, S.M.S and P.K proposed the growth mechanisms based on the experimental analysis. D.O, C.L and D.J.C performed the device fabrication, electrical measurements and related data analysis. Y.L assisted with the materials growth experiments. J.Q and M.C.S assisted with the TEM experiments. S.A.K and S.M.S prepared the manuscript with the contribution from all other authors. T.S.J, P.K and J.A supervised the project.     

\section{Competing financial interests}
The authors declare no competing financial interests.

\section{Supplementary Information}

The Supplementary Information is available at: \url{https://sid.erda.dk/share_redirect/ER343t9DQl}

\bibliography{ref}

\end{document}